\documentclass[sigconf]{acmart}
\AtBeginDocument{%
  }

\setcopyright{none}
\copyrightyear{2027}
\acmYear{2027}
\acmDOI{}
\acmConference[Under review]{Under review}
\acmISBN{}

\usepackage{booktabs}

\AtBeginDocument{%
  \captionsetup{
    font={it,md,small},
    labelfont={it,md},
    textfont={it,md},
    labelsep=period,
  }%
}

\begin{document}

\title{A Human-Augmenting Agentic Workflow for Observational Causal Inference}

\author{Winston Chou}
\authornote{Corresponding author.}
\email{wchou@netflix.com}
\orcid{0000-0001-7950-8015}
\affiliation{%
  \institution{Netflix}
  \city{New York}
  \state{NY}
  \country{USA}
}

\author{Adrien Alexandre}
\email{aalexandre@netflix.com}
\affiliation{%
  \institution{Netflix}
  \city{Seattle}
  \state{WA}
  \country{USA}
}

\author{Lars Olds}
\email{golds@netflix.com}
\affiliation{%
  \institution{Netflix}
  \city{Portland}
  \state{OR}
  \country{USA}
}

\author{Yi Zhang}
\email{yiz@netflix.com}
\affiliation{%
  \institution{Netflix}
  \city{Los Gatos}
  \state{CA}
  \country{USA}
}

\author{Nathan Kallus}
\email{nkallus@netflix.com}
\affiliation{%
  \institution{Netflix}
  \city{New York}
  \state{NY}
  \country{USA}
}
\additionalaffiliation{%
  \institution{Cornell University}
  \city{New York}
  \state{NY}
  \country{USA}
}

\renewcommand{\shortauthors}{Chou et al.}

\newcommand{\agent}{\texttt{oci-agent}}

\begin{abstract}
    Data analysis agents are becoming increasingly common tools for applied and scientific research. Yet, for highly specialized tasks such as Observational Causal Inference (OCI), human oversight remains necessary to ensure the validity of results. We introduce \agent, an open-source Python package that implements a human-in-the-loop agentic workflow for observational causal inference. \agent~is designed to automate vital but laborious aspects of applied causal inference, such as covariate balance checking, propensity score trimming, and sensitivity analysis, so that humans can focus on more nuanced tasks, such as framing questions, scrutinizing assumptions, and evaluating diagnostics and results.

    We initially open-sourced \agent~in June 2026 with support for doubly robust learning of the average treatment effect of a single binary treatment. Since then, we have added support for heterogeneous treatment effect estimation and for multiple continuous treatments via partially linear models. In this paper, we describe the principles behind \agent~and offer internal Netflix case studies and evaluations on public data of its capabilities. Across numerous evaluations, \agent~outperforms less structured baselines while remaining competitive with hand-tuned benchmarks. \agent~is used extensively for causal inference at Netflix and has orchestrated more than 100 analyses per month since its release in June.
\end{abstract}



\maketitle

\section{Introduction}

Data analysis is increasingly being delegated to artificial intelligence \cite{anthropic2026analytics,xu2026dataagent}. However, even for frontier models, human oversight and quality control remain essential.\footnote{For example, Anthropic requires humans to certify any analysis for leadership \cite{anthropic2026analytics}.} This is especially true for specialized tasks, such as Observational Causal Inference (OCI), which require substantial craft and domain expertise.

To enable trustworthy AI-assisted OCI, in both Netflix and the broader data science community, we introduce \agent, an open-source agentic workflow for OCI under unconfoundedness assumptions~\cite{imbens2004nonparametric,rubin1974estimating,rosenbaum1983propensity}. Our workflow is designed so that agents adhere to standardized and exhaustive templates for causal inference tasks. At the same time, \agent~seeks to empower \emph{human} inspection and evaluation.  To do this, it surfaces each step of the analysis through transparent and reproducible plans, specifications, diagnostics, and notebooks.

OCI requires context and care to perform well. However, parts of it can be repetitive and prone to error: checking and rechecking covariate balance, assessing propensity score overlap, running robustness checks, and tracking revisions. \agent~aims to reduce this toil so that OCI practitioners can focus on more nuanced tasks, such as framing questions, scrutinizing assumptions, and evaluating diagnostics and results.  To make these ideas accessible to the wider community, we have open-sourced a standalone version of \agent~for practitioners to use, extend, and improve.\footnote{\url{https://github.com/Netflix-Skunkworks/oci-agent}} This public package mirrors our internal Netflix repository, but only uses open-source libraries and evaluations on public data.

This paper makes four contributions. First, we open-source a deployed workflow that leverages AI for OCI, a highly specialized, judgment-heavy data analysis task. Second, we detail a human-in-the-loop pattern built around inspectable intermediate artifacts, such as reproducible notebooks, transparent diagnostics, and critical summaries. Third, we present Netflix case studies showing how our workflow improves OCI analyses in practice, for example, by detecting diagnostic failures, orchestrating robustness checks, replicating analyses in subgroups, and comparing related treatments. Fourth, we perform numerous evaluations on public data of \agent's capabilities.  We show that \agent~outperforms less structured baselines while remaining competitive with hand-tuned benchmarks, underscoring the necessity and leverage of our approach.

\section{Background}

\agent~is built on top of Netflix’s pre-LLM OCI toolkit \cite{lal2026estimating}. We developed this toolkit to answer ``point-in-time'' causal questions, such as ``what is the effect of playing a Netflix game on member retention?'' or ``what is the effect of watching a highly popular show on subsequent engagement?'' Questions of this kind inform business strategy, guide metric development, and contribute to a rich understanding of what drives member satisfaction.

Our toolkit is guided by a \emph{target trial emulation} philosophy~\cite{hernan2016target}. That is, for any point-in-time OCI question, we ask ``what is the ideal A/B test that would address this question?'' This A/B test may be expensive, slow, or even infeasible to run.  For example, we cannot randomize whether a member in fact plays a Netflix game or watches a highly popular show.  However, the thought exercise helps to pin down the assumptions needed for a credible answer, such as unconfoundedness of the treatment \cite{imbens2004nonparametric}.

To make the target trial analogy actionable, at the heart of our toolkit is a series of design diagnostics. For example, our \emph{covariate balance} diagnostic assesses whether the standardized mean difference of pre-treatment covariates between treatment and control groups is less than 0.2, possibly after reweighting \cite{austin2015moving}; and our \emph{overlap} diagnostic checks whether the estimated probability of treatment (aka propensity score) is bounded between 0.1 and 0.9 \cite{crump2009limited}.  These diagnostics assess whether a given causal analysis is drawing fair comparisons between treated and untreated units, or if there are hidden differences that could undermine its conclusions.

As we expand our OCI toolkit to agents, diagnostics and evaluation remain paramount. The standard approach to evaluating agents involves programmatically comparing their outputs to ground truth. Yet, outside of simulation and rare benchmark settings, the fundamental problem of causal inference implies that there is no ground truth in OCI \cite{holland1986statistics}.

Although this does not lessen the need for programmatic evaluations on synthetic benchmarks, which our workflow also embraces, one of our core principles is to empower \emph{human} evaluation by making each analytic step as transparent as possible. This principle is motivated by human-computer complementarity \cite{fuegener2026roles}: while computers excel at diligent execution of repetitive actions, such as applying a consistent set of rules or replicating an analysis with many different parameters, humans excel at applying rich contextual knowledge that may not exist in a model's training data, such as deciding which covariates to control for, whether an estimated effect size is meaningful, or if an estimate is stable enough to share with decision-makers.

To harness this complementarity, \agent~publishes intermediate artifacts in the form of plans, specifications, and notebooks that are easy for humans to inspect, edit, and re-execute. Thus, in the absence of actual ground truth, we rely on process audits and human oversight to build trustworthy agents.




\section{Implementation of \agent}

Our workflow has three key personas. First, the \emph{principal} is the human user (e.g., data scientist) who is ultimately accountable for conducting a thorough and correct analysis. Second, the \emph{actor} is the software persona that performs the analysis, including the design diagnostics. Third, the \emph{critic} is the software persona that synthesizes results, identifies gaps, and offers suggestions to improve the analysis. \agent~orchestrates the latter two personas in an actor-critic loop: specifying and triggering the analysis as the actor, then interpreting results and diagnosing flaws as the critic.

To help principals oversee and audit each analytic step, we provide them with templated Python notebooks that use vetted, non-agentic OCI tools based on open-source libraries such as EconML and DoubleML~\cite{econml,doubleml}. Although LLMs are rapidly improving their ability to perform data analysis, equipping them with deterministic tools achieves three goals. First, it is efficient: agents do not need to reimplement well-worn techniques that are already backed by robust libraries and open-source communities. Second, it makes analysis routines shorter and therefore easier for humans to review. Third, it ensures reproducibility and accountability: principals can re-execute the notebook to verify for themselves how a number was computed.

\begin{figure*}
    \centering
    \includegraphics[width=0.8\linewidth]{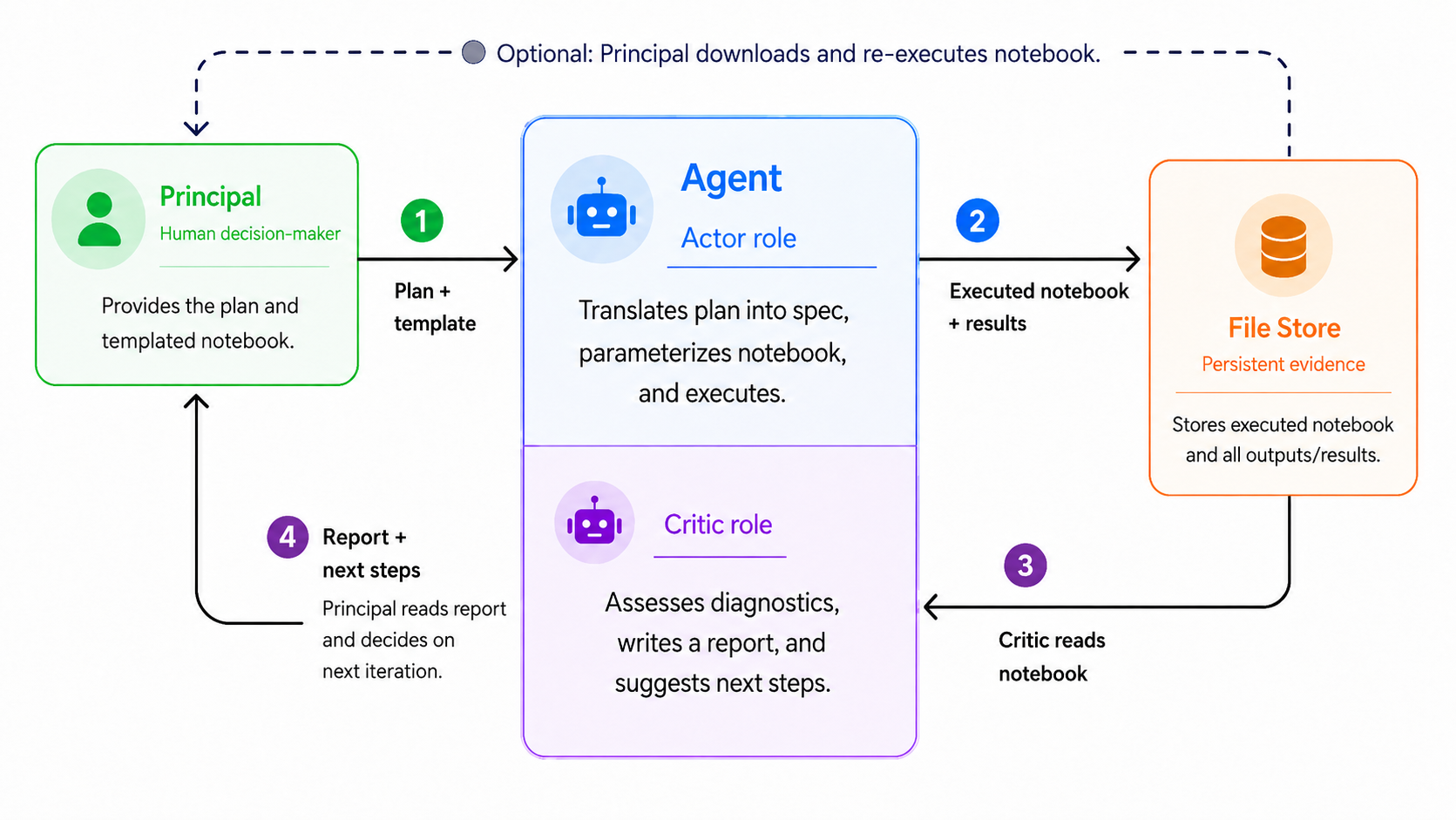}
    \begin{minipage}{.9\columnwidth}
    \caption{The \agent~workflow.}
    \Description{Diagram of the actor-critic workflow showing the principal providing an analysis plan to the actor, which produces a spec and executes a templated notebook with diagnostics, and the critic synthesizing results and reporting a credibility level back to the principal.}
    \label{fig:workflow}
    \end{minipage}
\end{figure*}

The principal's remaining responsibilities are to write the initial analysis plan and evaluate the analysis artifacts (i.e., the executed notebook and critic's report). To enable thorough evaluation, agents version-control their reports and upload executed notebooks to a file store, where they can be downloaded and re-executed by principals if they wish. We diagram this workflow in Figure~\ref{fig:workflow}.

\section{Deep Dives on Agent Capabilities}

The next three sections deep-dive three capabilities of \agent: doubly robust estimation of the treatment effect of a single binary treatment (Section~\ref{sec:binary}), estimating conditional average treatment effects within user segments (Section~\ref{sec:htes}), and estimating the effect of multiple and possibly continuous treatments via a partially linear model (Section~\ref{sec:plm}). The first capability accompanied our initial release of \agent~in June 2026~\cite{chou2026agentic}; since then we have added the second and third capabilities.

For each capability, we define the causal inference task, provide a Netflix case study, and describe the evaluations on public data implemented in the open-source package.

\section{Doubly Robust Estimation of the Effect of a Single Binary Treatment}
\label{sec:binary}

Our initial release of \agent~targeted the canonical OCI task: estimating the Average Treatment Effect (ATE) of a single binary treatment under unconfoundedness.\footnote{In addition to the ATE, \agent~is able to estimate the Average Treatment Effect on the Treated (ATT) population or the overlapping population for whom treatment can be considered ``as if'' randomly assigned (the Average Treatment Effect on the Overlap [ATO] population).} Given a treatment, an outcome, and a set of confounders, the templated notebook fits doubly robust Augmented Inverse Propensity Score Weighting (AIPW) estimators~\cite{robins1994estimation,bang2005doubly} of the ATE, ATT, and ATO, along with the design diagnostics described previously.

Let $T_i \in \{0,1\}$ denote treatment, $Y_i$ the outcome, and $X_i$ the pre-treatment covariates. The ATE is defined as $\tau = E[Y_i(1)-Y_i(0)]$, where $Y_i(t)$ denotes the potential outcome of observation $i$ under treatment value $t$. The actor estimates the propensity score $e(x)=P(T_i=1 \mid X_i=x)$ and outcome regressions $\mu_t(x)=E[Y_i \mid T_i=t, X_i=x]$ using cross-fitted nuisance models. It then computes the AIPW score
\begin{equation}
    \hat{\phi}_i =
    \hat{\mu}_1(X_i)-\hat{\mu}_0(X_i)
    + \frac{T_i(Y_i-\hat{\mu}_1(X_i))}{\hat{e}(X_i)}
    - \frac{(1-T_i)(Y_i-\hat{\mu}_0(X_i))}{1-\hat{e}(X_i)}
\end{equation}
and reports $\hat{\tau}=n^{-1}\sum_i \hat{\phi}_i$ with uncertainty estimates. Under the standard unconfoundedness assumptions, this estimator is doubly robust for the ATE, meaning that it is consistent when either $\mu_t$ or $e$ are consistent.  When the diagnostics imply insufficient overlap, the workflow trims observations with propensity scores outside pre-specified bounds~\cite{crump2009limited} and requires the critic to flag that the trustworthy estimand is the ATO rather than the ATE.

\subsection{Netflix Case Study: Estimating the Impact of New Content Engagement on Retention}

Here, we turn to a real-world example of \agent's deployment at Netflix. In recent years, Netflix has offered new entertainment types beyond streaming video to subscribers. A key business question is how these new entertainment types affect members' satisfaction and their likelihood of continuing to subscribe (aka retention).

To analyze the retention impact of one of these new entertainment types, which we will call Type X, we provided an analysis plan to \agent~that specifies our treatment (engagement with Type X during a pre-specified exposure window), outcome (retention two months after that window), and potential confounders, including pre-treatment Type X engagement and engagement with streaming video broadly.

To establish a baseline, we fed this analysis plan without additional scaffolding to Claude Sonnet 4.6, a powerful yet accessible general-purpose model. The model chose and executed a defensible analysis strategy --- linearly regressing retention on Type X engagement along with the controls --- and returned a polished report.  Yet, when we ran the same analysis through our paved-path tooling and agentic workflow, also using Sonnet 4.6, \agent~produced an estimate that was just 25\% of the baseline. What explains the difference between the baseline and the paved-path estimates?

A core challenge when analyzing new entertainment types is \emph{early adopter bias}: the first users of any new offering are likely to be systematically different from the general population. For example, they may be heavier users of Netflix generally, or they may be extremely strong fans of the underlying titles. Early adopter bias manifested in our analysis as poor overlap of the estimated propensity scores: the vast majority of observations had a minuscule estimated probability of engaging with Type X. Our critic persona flagged this issue in its review of the actor's analysis.

\begin{figure}
    \centering
    \includegraphics[width=\linewidth]{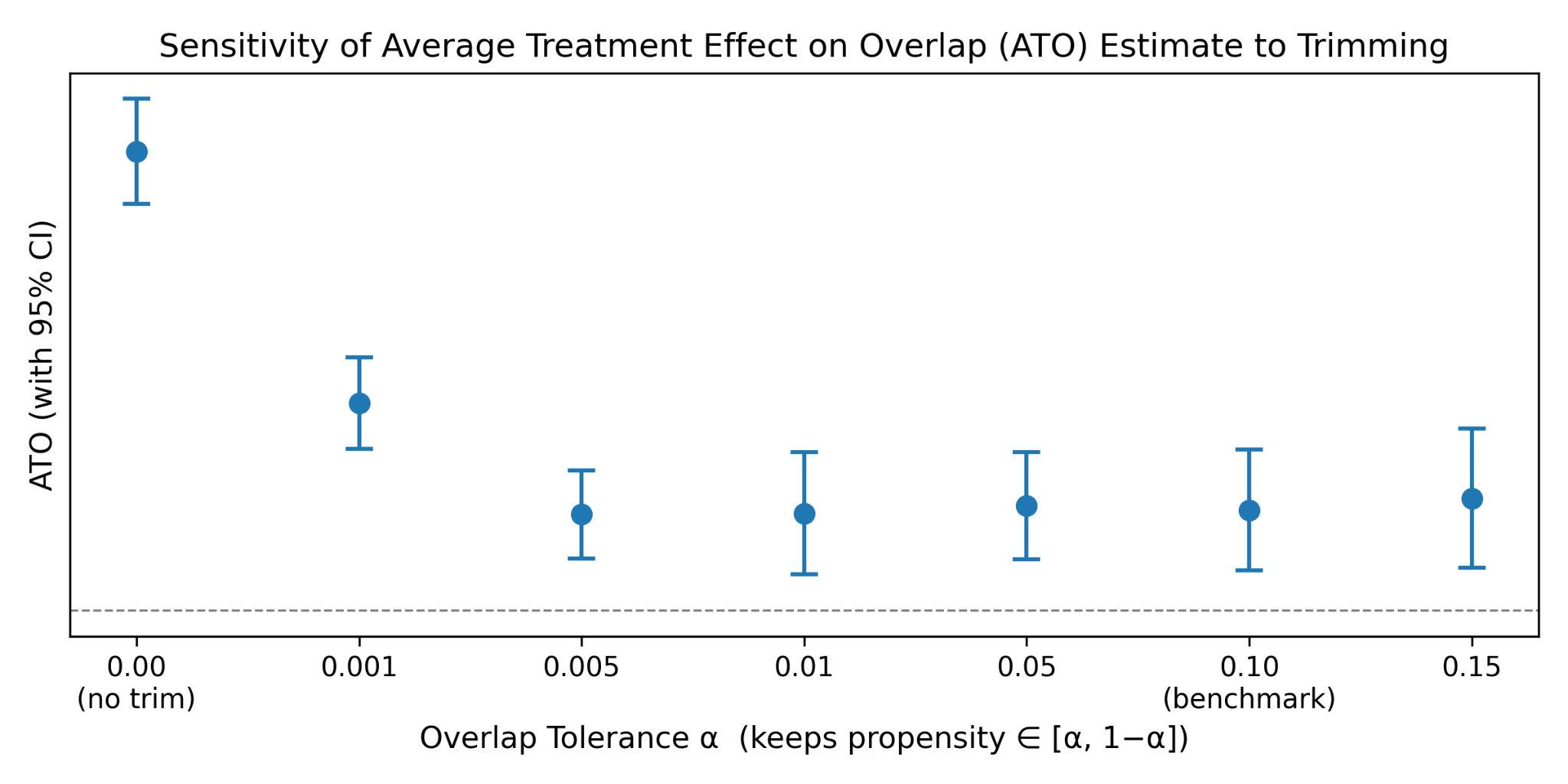}
    \begin{minipage}{.9\columnwidth}
    \caption{Estimated ATE on the overlapping population versus trimming threshold; the ATO is stable within [0.005, 0.995].}
    \Description{Line plot showing the estimated ATE on the overlapping population as a function of the propensity-score trimming threshold, remaining approximately flat within bounds of 0.005 to 0.995.}
    \label{fig:sensitivity}
    \end{minipage}
\end{figure}

To address this and other diagnostic failures, our workflow gives agents a playbook. For example, to overcome poor overlap, we instruct the agent to trim units with estimated propensity scores outside the range [0.1, 0.9]. This scopes the treatment effect being estimated to the average treatment effect in the ``overlap'' population (ATO) that is not extremely likely or unlikely to engage in the new entertainment type, an important caveat that we instruct the critic to flag in its report. One benefit of delegating trimming to \agent~is that it is very easy to assess the sensitivity of the analysis to the choice of range, as demonstrated in Figure~\ref{fig:sensitivity}.  As the figure shows, the estimated ATO is quite stable after trimming units with propensities of Type X engagement outside the range [0.005, 0.995].

\subsection{Evaluations: ACIC 2016 and LaLonde}

Here, we share evaluations on public data of \agent's ability to estimate the causal effect of a binary treatment.  Our initial open-source release of \agent~contained evaluations on synthetic datasets from the 2016 Atlantic Causal Inference Conference (ACIC) Data Competition~\cite{dorie2019automated}. Since then, we have also added evaluations on non-synthetic data based on the classic LaLonde study~\cite{lalonde1986evaluating}.

\subsubsection{Accuracy against ACIC 2016 benchmarks}

Our first evaluation runs our binary treatment notebook on 231 datasets generated by randomly sampling three datasets for each of the 77 data-generating processes (DGPs) in the ACIC data. Next, it uses the critic to grade the resulting estimates as either satisfactory or unsatisfactory based on the diagnostics.

\begin{figure}
    \centering
    \includegraphics[width=\linewidth]{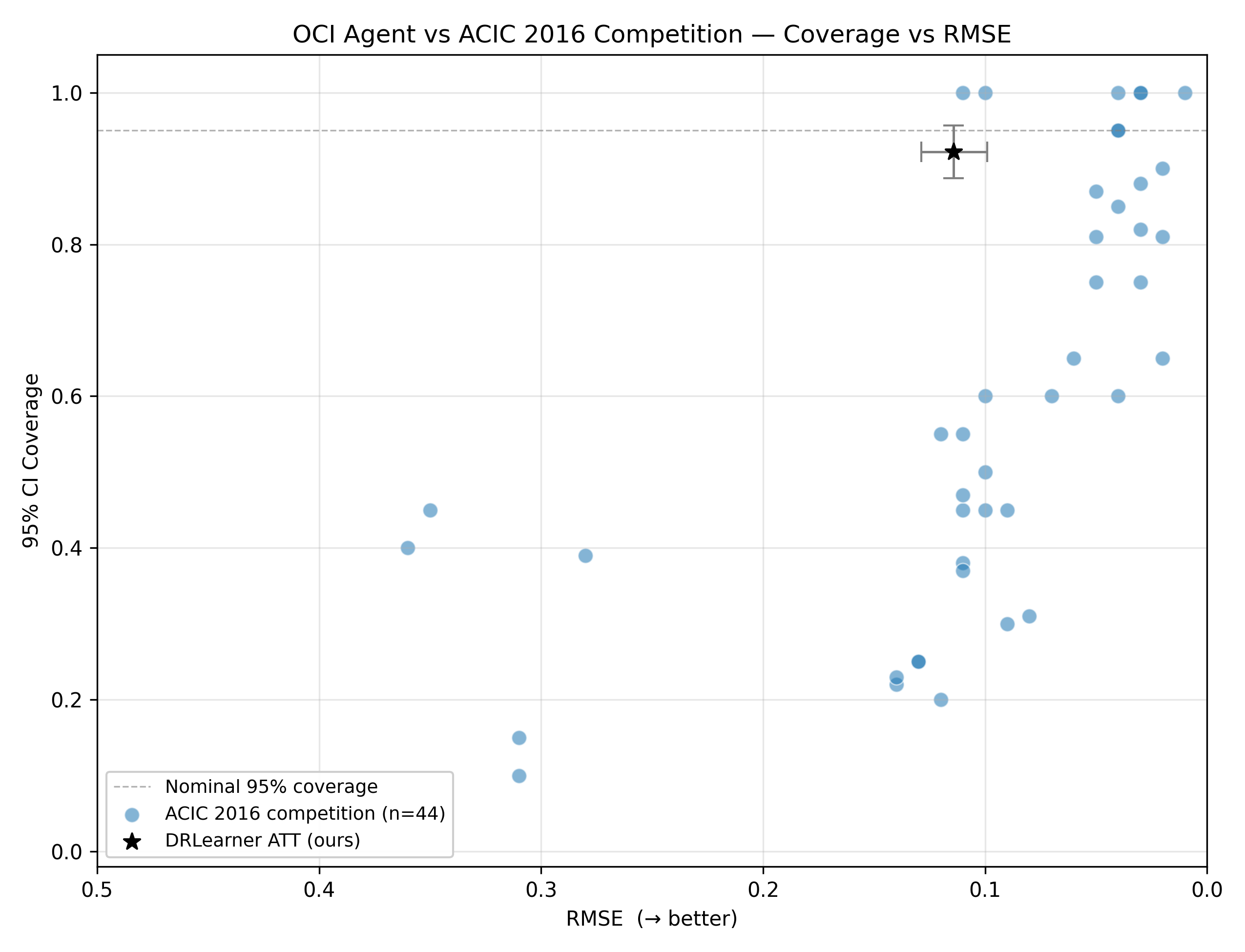}
    \begin{minipage}{.9\columnwidth}
    \caption{RMSE and 95\% CI coverage of \agent's ATT estimates versus the 44 ACIC 2016 competitor methods.}
    \Description{Scatterplot of RMSE versus confidence-interval coverage for the ACIC competitor methods, with the oci-agent method achieving low RMSE and coverage near 95 percent.}
    \label{fig:benchmark}
    \end{minipage}
\end{figure}

Figure~\ref{fig:benchmark} plots the average RMSE and coverage of 95\% confidence intervals of our ATT estimates against the 44 competitor methods in the ACIC competition. As the scatterplot shows, our statistical methodology is competitive against these benchmarks, many of them hand-tuned: it achieves low RMSE and well-calibrated confidence intervals that cover the truth in approximately 95\% of datasets.

\subsubsection{Certifying reliable estimates}

\begin{figure}
    \centering
    \includegraphics[width=\linewidth]{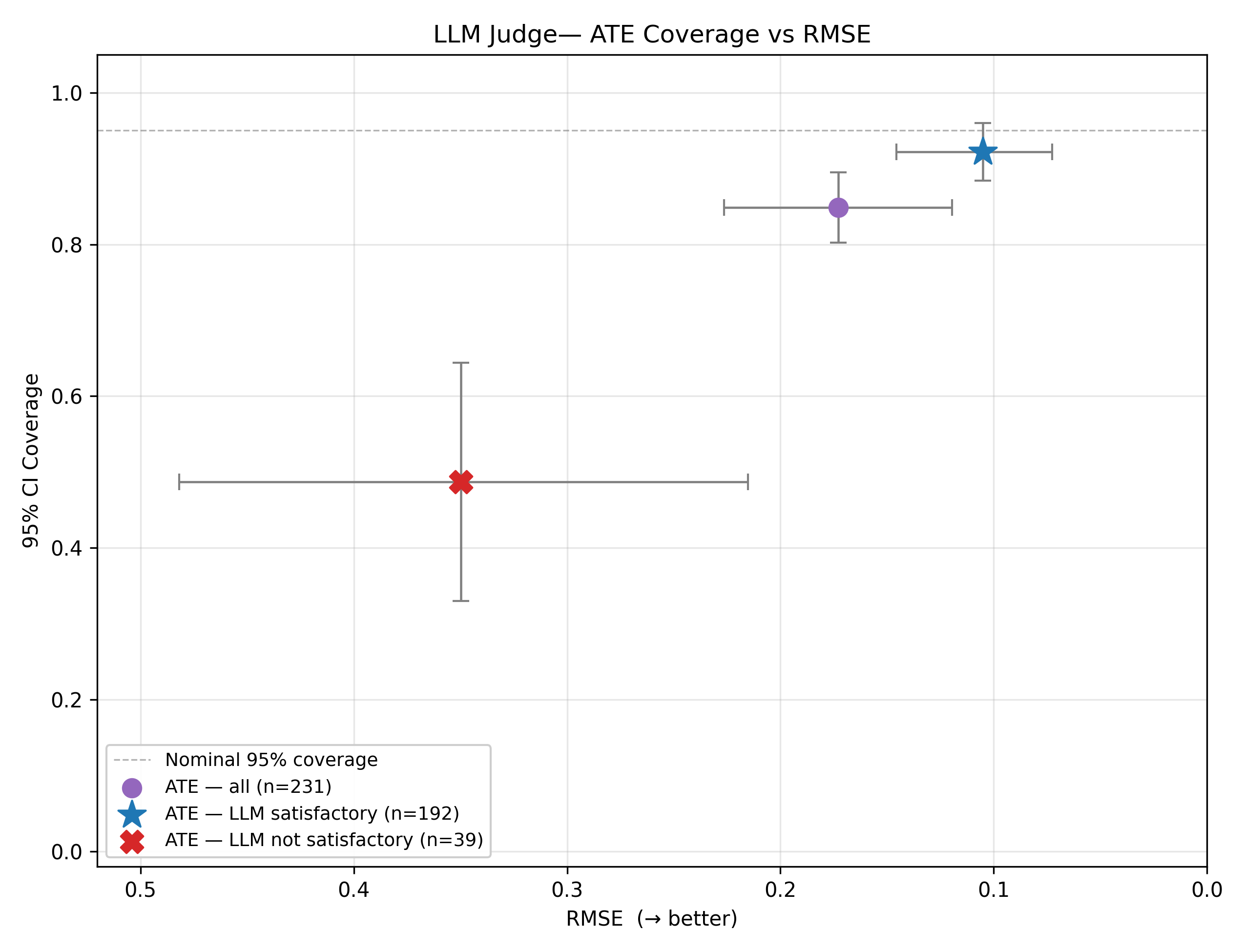}
    \begin{minipage}{.9\columnwidth}
    \caption{ATE estimates by critic grade: all 231 (purple), 192 satisfactory (blue), 39 unsatisfactory (red).}
    \Description{Scatterplot of RMSE versus coverage with three markers: all estimates, satisfactory estimates, and unsatisfactory estimates. Satisfactory estimates cluster at lower RMSE and coverage near 95 percent, while unsatisfactory estimates have higher RMSE and poorer coverage.}
    \label{fig:judge}
    \end{minipage}
\end{figure}

More to the point, our diagnostics and agentic workflow help distinguish more reliable estimates from less reliable estimates. To illustrate this, Figure~\ref{fig:judge} plots our ATE estimates in terms of RMSE and coverage.\footnote{Note that the ATT is largely identifiable in these datasets while the ATE is not~\cite{dorie2019automated}.  Therefore, this eval focuses on the ATE in order to have a meaningful number of failing diagnostics.} We separately plot the RMSE and coverage of all 231 estimates (purple dot), the 192 estimates judged to be satisfactory (blue star), and the 39 estimates judged to be unsatisfactory (red dot).  As the figure shows, when equipped with our diagnostic suite, the critic agent is consistently able to separate reliable estimates from unreliable estimates: the satisfactory estimates have significantly lower RMSE and better calibrated confidence intervals than the unsatisfactory estimates.

\subsubsection{Scaffolding versus pure prompting}

\begin{figure}
    \centering
    \includegraphics[width=\linewidth]{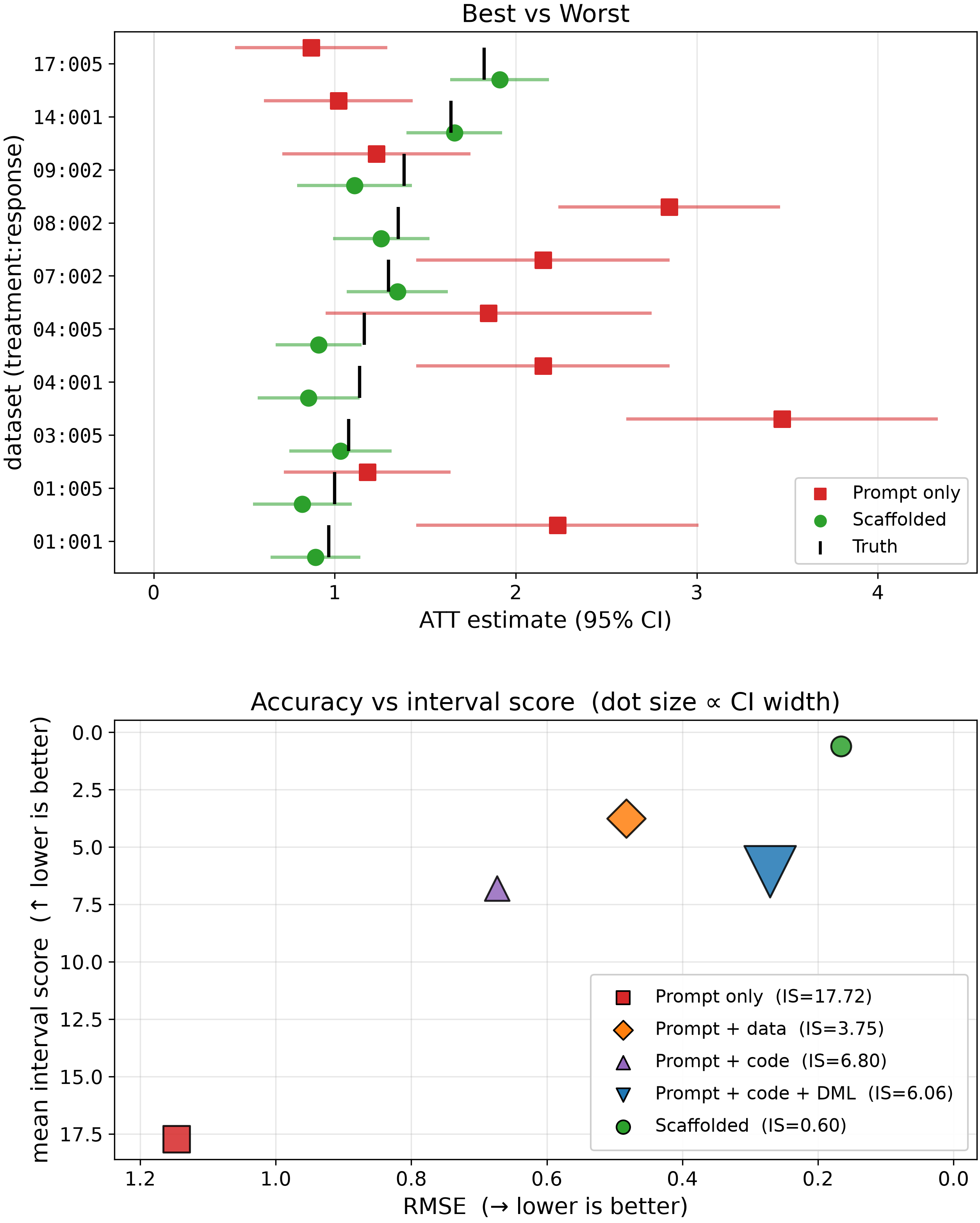}
    \begin{minipage}{.9\columnwidth}
    \caption{Capability-ladder ablation on ten ACIC 2016 datasets. Top: best vs.\ worst (prompt-only vs.\ scaffolded) against ground truth. Bottom: accuracy (RMSE) versus mean interval score for all five arms (dot size proportional to mean CI width); the scaffolded agent is both the most accurate and the best-calibrated.}
    \Description{A stacked figure. The top panel is a forest plot comparing prompt-only and scaffolded estimates against ground truth for ten ACIC datasets, with scaffolded tracking the truth closely and prompt-only not. The bottom panel is a scatterplot of RMSE versus mean interval score for all five capability-ladder arms, with the scaffolded agent in the most accurate and best-calibrated corner and prompt-only worst.}
    \label{fig:baseline-scaffolded}
    \end{minipage}
\end{figure}

Our most direct evaluation fixes the analysis plan and varies the capabilities we give the LLM (Claude Sonnet 4.6). The results of this evaluation are shown in Figure~\ref{fig:baseline-scaffolded} for ten randomly-selected ACIC 2016 datasets.  We compare five layers of scaffolding: (i) \emph{prompt only}, which only gives the model the plan, a small sample of the dataset, and a path to the full data file; (ii) \emph{prompt + data}, which pastes the full realized dataset into the prompt; (iii) \emph{prompt + code}, which lets the model write and execute Python with an estimator of its choosing; (iv) \emph{prompt + code + DML}, which further instructs it to use double machine learning~\cite{chernozhukov2018double}; and (v) \emph{scaffolded}, the full \agent~notebook workflow. 

The top panel of Figure~\ref{fig:baseline-scaffolded} demonstrates that \agent~radically outperforms the least capable tier (prompt only) across all datasets. Given only a prompt, the model hallucinates responses that are wholly uncorrelated with ground truth. Equipped with our scaffolding, the \emph{same} LLM recovers the ground truth in nine out of ten datasets, with estimates that correlate strongly with ground truth.  The bottom panel compares all five layers simultaneously on accuracy and calibration by plotting each layer's RMSE against its mean \emph{interval score}~\cite{winkler1972decision,gneiting2007strictly}, a proper scoring rule that penalizes both undercoverage and excessively wide confidence intervals. \agent~dominates all layers on both axes and is simultaneously the most accurate and the best calibrated.

\subsubsection{Evaluations on LaLonde}

Thus far, the evaluations use synthetic data for which ground truth is known. To test \agent~on real data with an externally validated target, we turn to the classic job training study by LaLonde~\cite{lalonde1986evaluating} on the National Supported Work Demonstration experiment.  This experiment estimated an average effect of job training on 1978 earnings of \$1{,}794. Following Dehejia and Wahba~\cite{dehejia2002propensity}, we discard the experimental controls and substitute the two standard non-experimental comparison groups sampled from the PSID and CPS surveys. These groups overlap poorly with the treated units, so a naive difference in means is badly biased ($-$\$15{,}205 for PSID and $-$\$8{,}498 for CPS). Recovering the experimental estimate of \$1{,}794 requires correctly adjusting for confounding under weak overlap. We give the agent only the observational data and score the absolute percentage error of its ATT estimate against the experimental target of \$1{,}794.

\begin{figure}
    \centering
    \includegraphics[width=\linewidth]{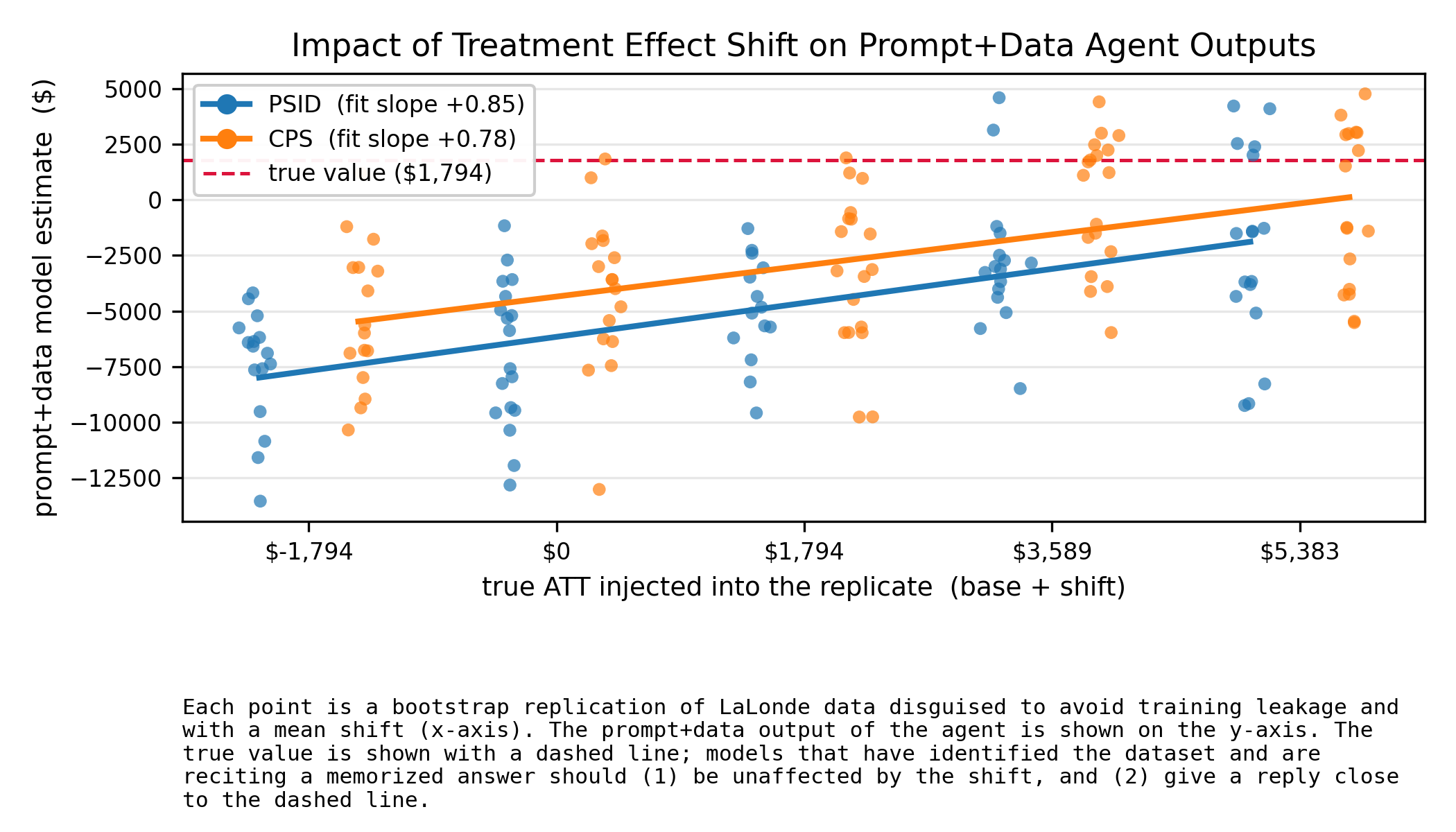}
    \begin{minipage}{.9\columnwidth}
    \caption{Leakage test on the disguised LaLonde data (\emph{prompt + data} arm): the
    estimated ATT tracks the shift injected into the true effect (per-group OLS slope
    $\approx 0.8$) instead of pinning at the memorized \$1{,}794 (dashed line), for the
    PSID and CPS control groups.}
    \Description{Scatterplot of the model's estimated ATT versus the injected true
    effect for the PSID and CPS control groups, with per-group least-squares fit lines
    of slope near 0.8 and a dashed reference line at the memorized value of \$1{,}794.}
    \label{fig:lalonde-shift}
    \end{minipage}
\end{figure}

Because the LaLonde effect is widely reported in the literature, it may appear in the underlying model's training data, so we must rule out that the agent simply recites \$1{,}794 rather than estimating it. To do this, we run a leakage test: we ``disguise'' the LaLonde data by bootstrapping it, replacing the column names, and injecting a shift into the treated outcomes so that the true effect moves away from \$1{,}794 (by $-2\times$ to $+2\times$ the ATT). We then test whether the estimated effects correlate with this shift. A model reciting the memorized value would stay pinned near \$1{,}794 regardless of the shift, whereas a model that actually analyzes the bootstrapped data should give answers that correlate with it.  Figure~\ref{fig:lalonde-shift} demonstrates that, as long as we provide the model with minimal scaffolding (the \emph{prompt + data} tier from the above ablation ladder), the agent's estimates track the injected shift (OLS
slope $\approx 0.8$) and almost never land very close to \$1{,}794.\footnote{If we give the agent only the prompt, responses are weakly correlated with the shift but do not center on the true value, indicating hallucination rather than recitation.}  Thus, disguising the task prevents the model from reciting a memorized value.

\begin{figure}
    \centering
    \includegraphics[width=\linewidth]{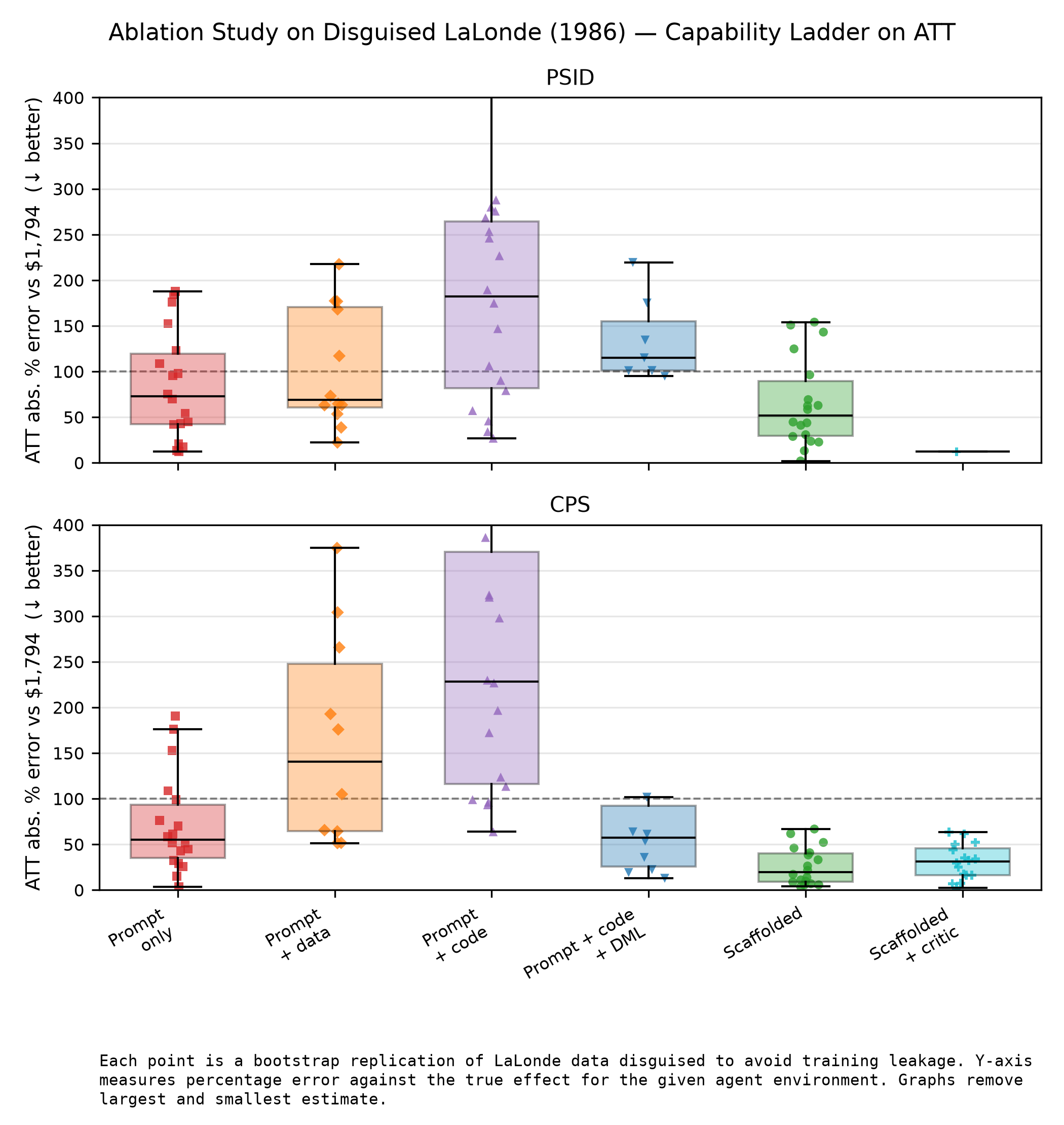}
    \begin{minipage}{.9\columnwidth}
    \caption{Disguised-LaLonde ablation: ATT absolute \% error versus the experimental
    target across the capability ladder, for the PSID (top) and CPS (bottom) control
    groups.}
    \Description{Two stacked box-plot panels of ATT absolute percentage error across
    the capability-ladder arms, one for the PSID control group and one for the CPS
    control group.}
    \label{fig:lalonde-ablation}
    \end{minipage}
\end{figure}

Having established that the LLM is not simply reciting the true ATT, we conduct the above ablation analysis on the disguised LaLonde task. Figure~\ref{fig:lalonde-ablation} reports the absolute percentage error of each ATT estimate relative to the experimental target across 20 bootstrap replications, separately for the PSID (top) and CPS (bottom) comparison groups. As the figure shows, only the fully scaffolded \agent~achieves reasonable accuracy.  Even letting the model write and run its own analysis code, with or without an explicit instruction to use double machine learning, does not reliably achieve error that is less in magnitude than the true ATT. In contrast, \agent~achieves median absolute percentage error of 52\% on PSID (the harder task) and 19\% on CPS. Note that, when we employ the critic persona, it withholds certification of the PSID task in all but one replication due to very poor overlap~\cite[cf.][]{dehejia2002propensity}.

\section{Heterogeneous Treatment Effects}
\label{sec:htes}

In this section, we turn to another common OCI task, which is to assess how treatment effects vary across user segments, for example, by country or baseline engagement levels. \agent~estimates heterogeneous treatment effects (HTEs) by reusing the doubly robust AIPW scores $\hat{\phi}_i$ estimated in the binary-treatment workflow. That is, letting $S_i$ be a categorical pre-treatment covariate, the actor estimates Conditional Average Treatment Effects (CATEs) by taking the average of $\hat{\phi}_i$ within each level of $S_i$.  Since most of the effort of identifying and estimating treatment effects lies in constructing $\hat{\phi}_i$, this makes it simple and efficient to compute CATEs after main effects are estimated \cite{seyfarth2025heterogeneous}.

\subsection{Netflix Case Study: Learning For Whom Treatments Matter}

Here, we share an internal case study of HTE estimation using \agent. In this study, the principal wanted to know whether the treatment had the same retention effect across member segments. As before, we constructed the treatment by defining an exposure window and the outcome by measuring retention two months after that exposure window.  We included numerous controls, including pre-treatment engagement levels and estimated churn probability.  Although the treatment in this instance was continuous rather than binary, we discretized it into five ordinal levels to enable dose-response estimation using AIPW \cite{lal2026estimating}.

\begin{figure}
    \centering
    \includegraphics[width=\linewidth]{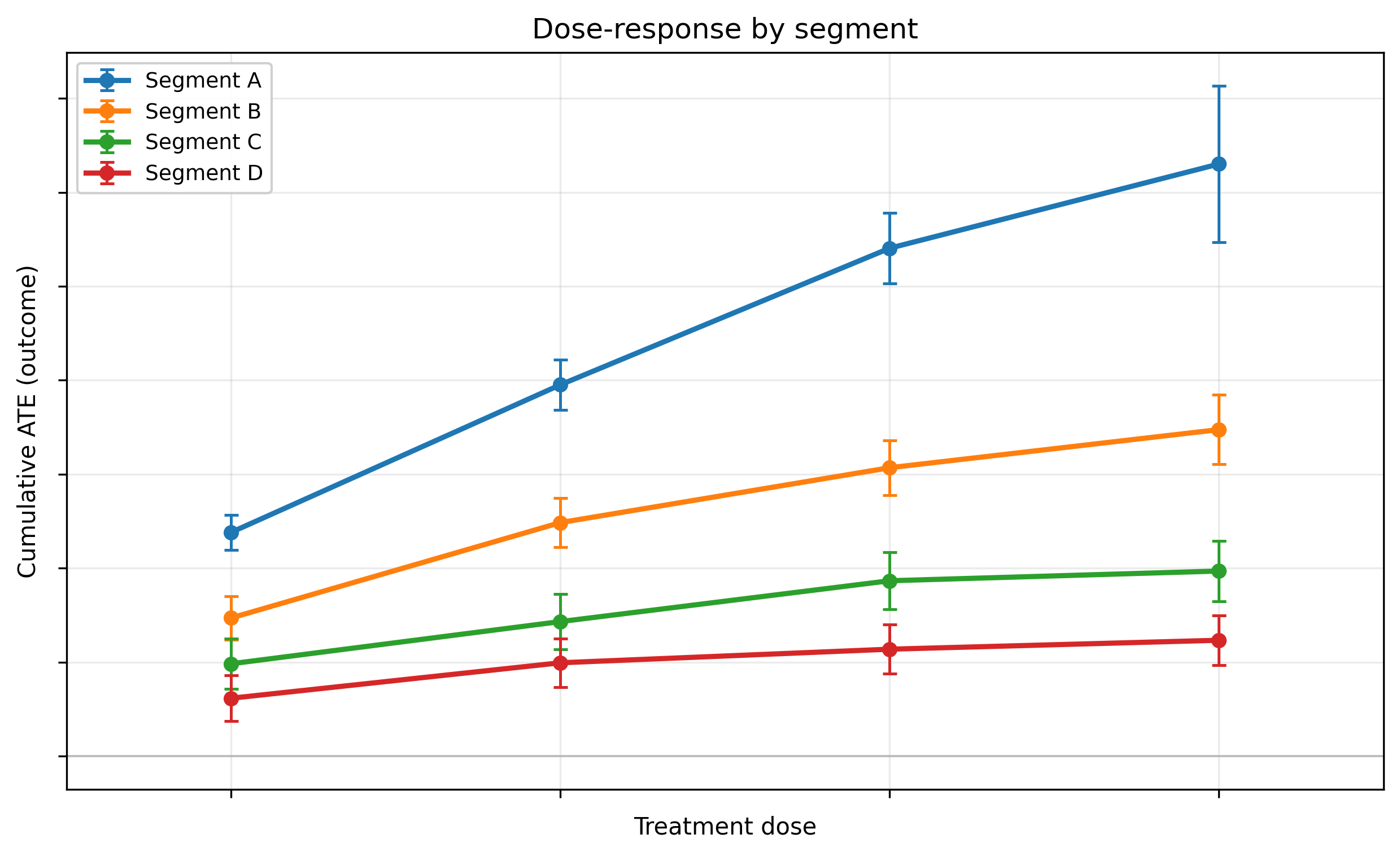}
    \begin{minipage}{.9\columnwidth}
    \caption{Internal HTE case study (anonymized): cumulative treatment effect by treatment dose and member segment.}
    \Description{Line plot of cumulative average treatment effects by treatment dose for four anonymized subscriber segments. Segment A has the steepest positive dose response, Segment B has a moderate positive response, Segment C has a small positive response, and Segment D is close to flat.}
    \label{fig:hte-casestudy}
    \end{minipage}
\end{figure}

While the estimated ATE was positive, the principal suspected it could hide important variation. To investigate this hypothesis, we used \agent's HTE functionality to estimate a separate dose-response curve for each of four member segments, labeled A through D in Figure~\ref{fig:hte-casestudy}.  As the figure shows, the dose-response curve varies significantly across segments, with Segment A benefiting the most from treatment and Segment D showing almost no incremental benefit at higher dosages.

Insights such as these help teams develop personalized experiences that better meet the needs of different member segments.  Because requests for HTE estimation are both common and easy to implement after ATE estimation, Netflix's OCI tooling displays visualizations akin to Figure~\ref{fig:hte-casestudy} by default \cite[][]{seyfarth2025heterogeneous}.

\subsection{Evaluations for Heterogeneous Treatment Effect Estimation}

\begin{figure}
    \centering
    \includegraphics[width=\linewidth]{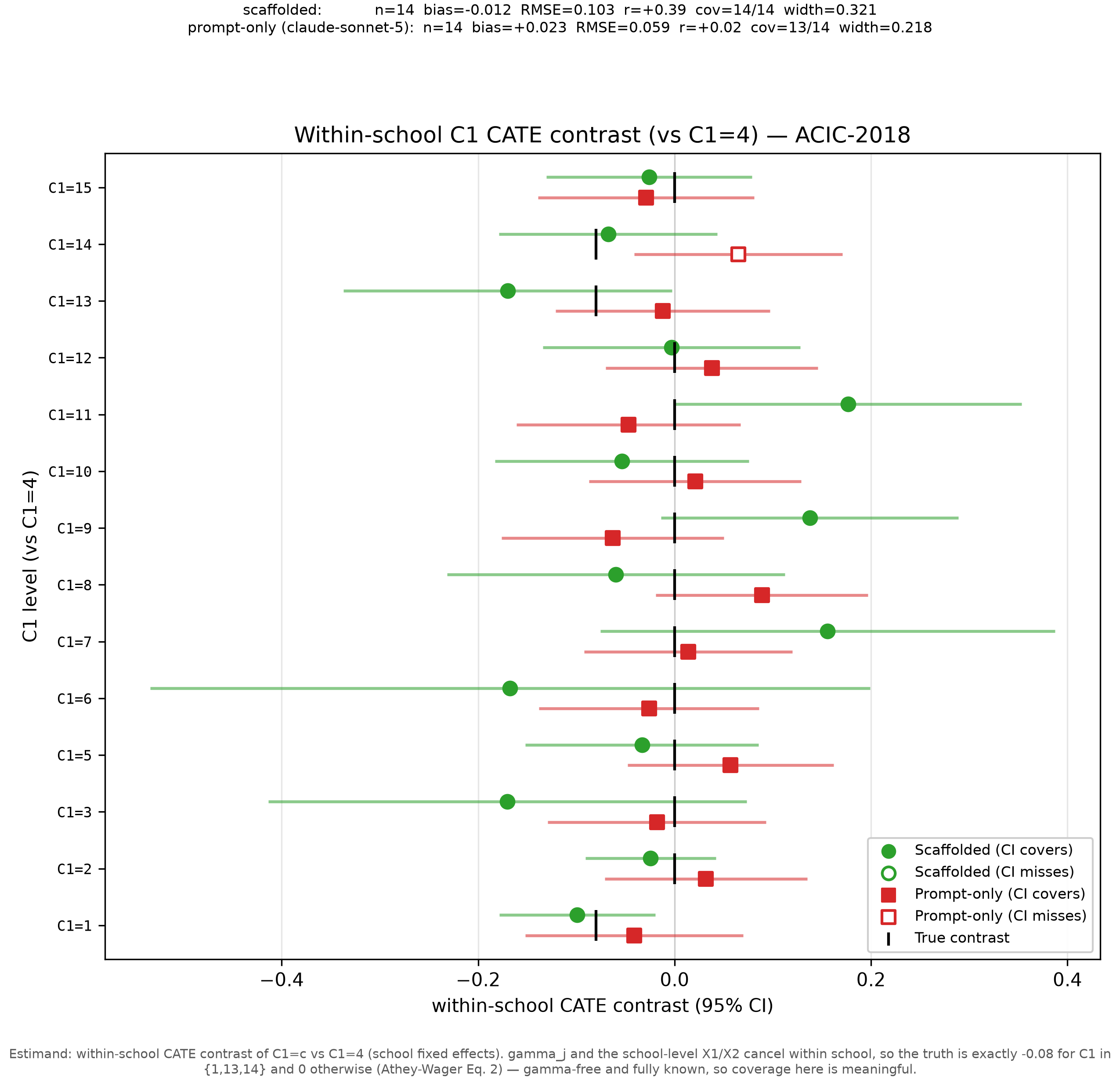}
    \begin{minipage}{.9\columnwidth}
    \caption{HTE on the ACIC 2018 data (best vs.\ worst). The near-zero, low-SNR contrasts make this an unrealistic benchmark: prompt-only is competitive only because guessing near zero is hard to beat.}
    \Description{Forest plot of within-group contrast estimates for the ACIC 2018 data comparing the prompt-only baseline and the scaffolded agent against ground truth, where the true contrasts are close to zero.}
    \label{fig:hte-acic2018}
    \end{minipage}
\end{figure}

To evaluate \agent's HTE functionality, we started with a public benchmark. A data challenge at ACIC 2018 featured a synthetic version of the National Study of Learning Mindsets experiment, which randomized a ``growth mindset'' educational program to students within schools \cite{carvalho2019assessing}. The organizers of that data competition injected HTEs using a complex functional form; unfortunately, this includes synthetic school random effects, which we were unable to locate or reproduce. Therefore, we were only able to evaluate \agent~on one dimension of heterogeneity: the organizers added a fixed negative offset to levels 1, 13, and 14 of the synthetically generated covariate $C_1$. As Figure~\ref{fig:hte-acic2018} shows, \agent~covers the CATE for every level of $C_1$, with the strongest negative estimates appearing on shifted levels such as $C_1=1$ and $C_1=13$. However, the prompt-only baseline looks deceptively competitive here, as it is ``correctly'' guessing estimates around 0.

To set a higher bar, we constructed our own synthetic DGP, which can be found in the public \agent~repository, which mirrors the structure of the ACIC 2018 dataset. Specifically, we simulate $G$ groups (``schools''), each with a random intercept $\gamma_j$ and group-level covariates that are constant within the group. Within each group, units carry a categorical covariate $C_1$ that varies across units, together with individual-level covariates. The treatment effect only depends on $C_1$, so that $\tau_i = \tau_0 + \alpha(C_1)$. Because $\gamma_j$ and the group-level covariates are constant within schools, they cancel when taking within-group differences.

We also add confounding: treatment assignment depends on both the individual covariates and the group effect $\gamma_j$, and the outcome depends on the same covariates plus a group-varying baseline. Thus, the difference-in-means within groups \emph{and} pooled covariate adjustment are biased; only an estimator that adjusts for the covariates and differences out the group effect recovers the true CATE.

\begin{figure}
    \centering
    \includegraphics[width=\linewidth]{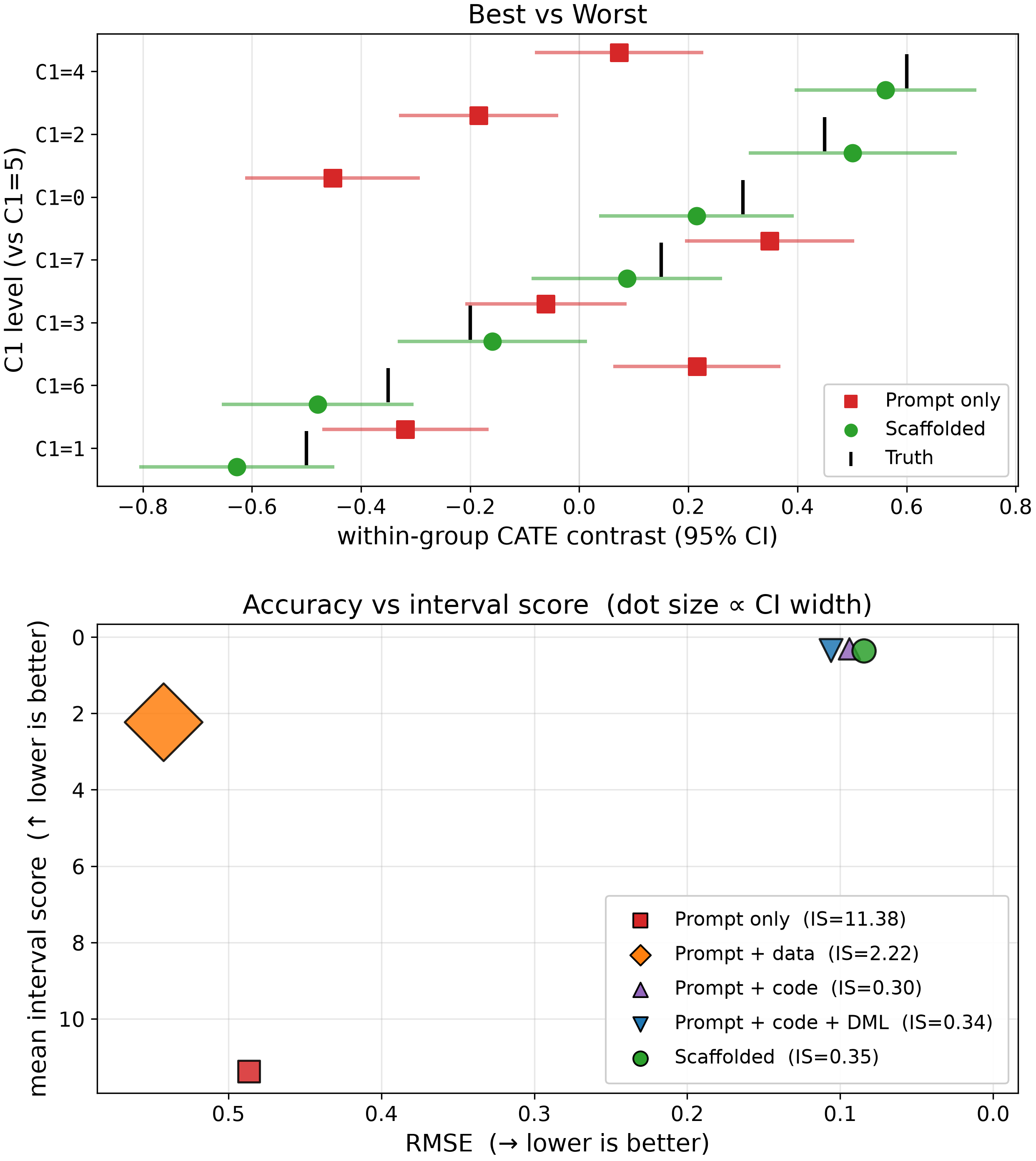}
    \begin{minipage}{.9\columnwidth}
    \caption{HTE on our synthetic within-group-contrast dataset. Top: best vs.\ worst (prompt-only vs.\ scaffolded) against the known contrasts. Bottom: accuracy (RMSE) versus mean interval score across all five arms (dot size proportional to mean CI width). The scaffolded agent recovers the strong, known contrasts while pure prompting does not.}
    \Description{A stacked figure for the synthetic HTE dataset. The top panel is a forest plot comparing prompt-only and scaffolded within-group contrast estimates against ground truth, with scaffolded tracking the true contrasts and prompt-only not. The bottom panel is a scatterplot of RMSE versus mean interval score for all five capability-ladder arms, with the scaffolded agent most accurate and best-calibrated and prompt-only worst.}
    \label{fig:hte-synthetic}
    \end{minipage}
\end{figure}

In Figure~\ref{fig:hte-synthetic}, the top panel shows that pure prompting results in highly biased estimates, whereas the fully scaffolded agent recovers the ground truth; the bottom panel confirms that the scaffolded agent also attains the best accuracy and interval score across the full capability ladder. Again, for reproducibility, we include this evaluation in the open-source \agent~repository.

\section{Multiple Continuous Treatments Using Partially Linear Models}
\label{sec:plm}
Thus far, our focus has been on individual binary and/or discrete treatments.  However, teams at Netflix are often interested in the effects of multiple simultaneous treatments.  For example, we may be interested in computing the ``exchange rate'' between Type X engagement and streaming video engagement.  Additionally, these treatments are often continuous (or at least take on more than two distinct values).

A useful way to model multiple continuous treatments is with a \emph{partially linear model} (PLM):
\begin{eqnarray}
    Y_i = A_i^\top\theta + g_0(X_i) + \zeta_i, &\qquad& E[\zeta_i | A_i, X_i] = 0 \label{eqn:plm} \\ \nonumber
    A_i = m_0(X_i) + \upsilon_i, &\qquad& E[\upsilon_i | X_i] = 0,
\end{eqnarray}
where $Y_i$ is the outcome, $A_i$ is a vector (possibly of length 1) of treatments, $\theta$ is the target parameter, $g_0$ and $m_0$ are nuisance functions, and $\zeta_i$ and $\upsilon_i$ are stochastic errors. \agent~supports PLM estimation (including partially linear logistic regression models) via the open-source DoubleML package~\cite{doubleml}.

Operationally, the actor estimates nuisance functions for the conditional outcome and conditional treatment means, residualizes both $Y_i$ and $A_i$ using cross-fitting, and solves the orthogonal moment condition for $\theta$~\cite{chernozhukov2018double}. This gives the principal a coefficient vector that can be interpreted as the marginal effect of each treatment on the outcome after adjusting flexibly for $X_i$, subject to the PLM assumptions.

\subsection{Netflix Case Study: Comparing the Effects of Multiple Treatments}

In this case study, the principal sought to compare the effects of two related treatments on retention. Treatment A is a continuous variable, while Treatment B is a sparse binary exposure. The key business question is whether each treatment has an independent effect on the otucome after controlling for the other (i.e., whether Treatment B is incremental ``on top'' of Treatment A and vice versa).

\begin{figure}
    \centering
    \includegraphics[width=\linewidth]{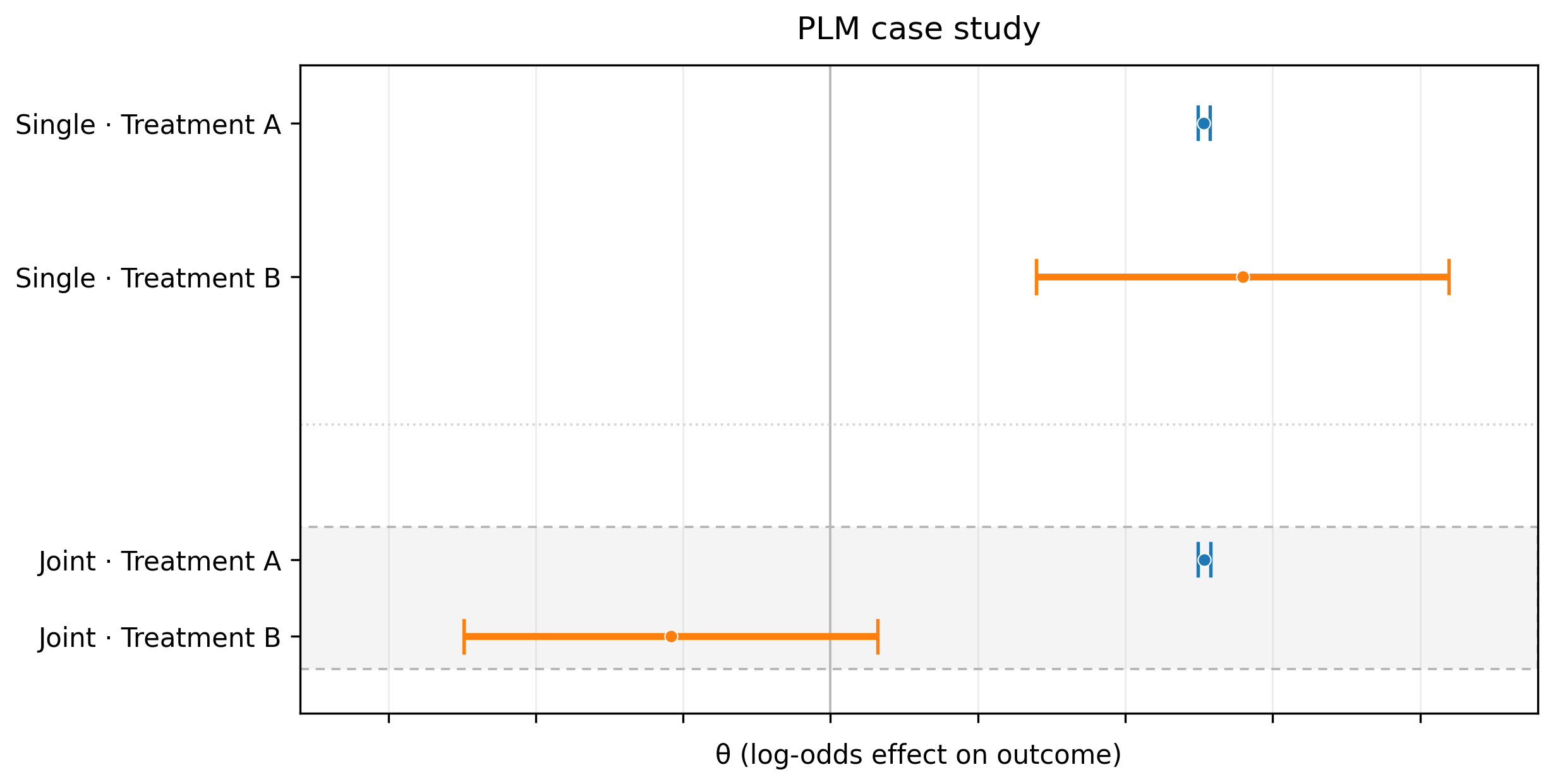}
    \begin{minipage}{.9\columnwidth}
    \caption{Internal PLM case study (anonymized): treatment coefficients estimated singly and jointly.}
    \Description{Forest plot of partially-linear-model coefficients for Treatment A and Treatment B. Treatment A is positive and stable in both the single-treatment and joint-treatment models, while Treatment B is positive alone but attenuates toward zero with wider uncertainty in the joint model.}
    \label{fig:plm-casestudy}
    \end{minipage}
\end{figure}

Our playbook for PLM runs in two modes. First, the actor estimates single-treatment models for Treatment A and Treatment B separately. Second, it estimates a joint model containing both treatments, using the same pre-treatment controls and cross-fitted nuisance models. Figure~\ref{fig:plm-casestudy} shows the resulting coefficients on the log-odds scale. The effect of Treatment A is positive and tightly estimated in both the single-treatment and joint-treatment specifications. The effect of Treatment B is significantly positive when estimated alone; however, it attenuates toward insignificance once Treatment A is included.  Treatment B also has much wider confidence intervals due to its sparsity.

These findings are of direct business relevance.  A single-treatment analysis would indicate that both treatments have roughly comparable effects. Analyzing them jointly reveals that part of Treatment B's effect likely reflects its correlation with Treatment A, and that the independent contribution of Treatment B is relatively uncertain.

\subsection{PLM Evaluations}

While we did not find an ideal public benchmark for this multi-treatment setting, we provide our own synthetic DGP, which supports continuous and binary treatments and outcomes, in the public repository. In the continuous outcome case, this DGP generates data from the PLM \eqref{eqn:plm}; in the binary outcome case, it generates data from a partially linear logistic model.

\begin{figure}
    \centering
    \includegraphics[width=\linewidth]{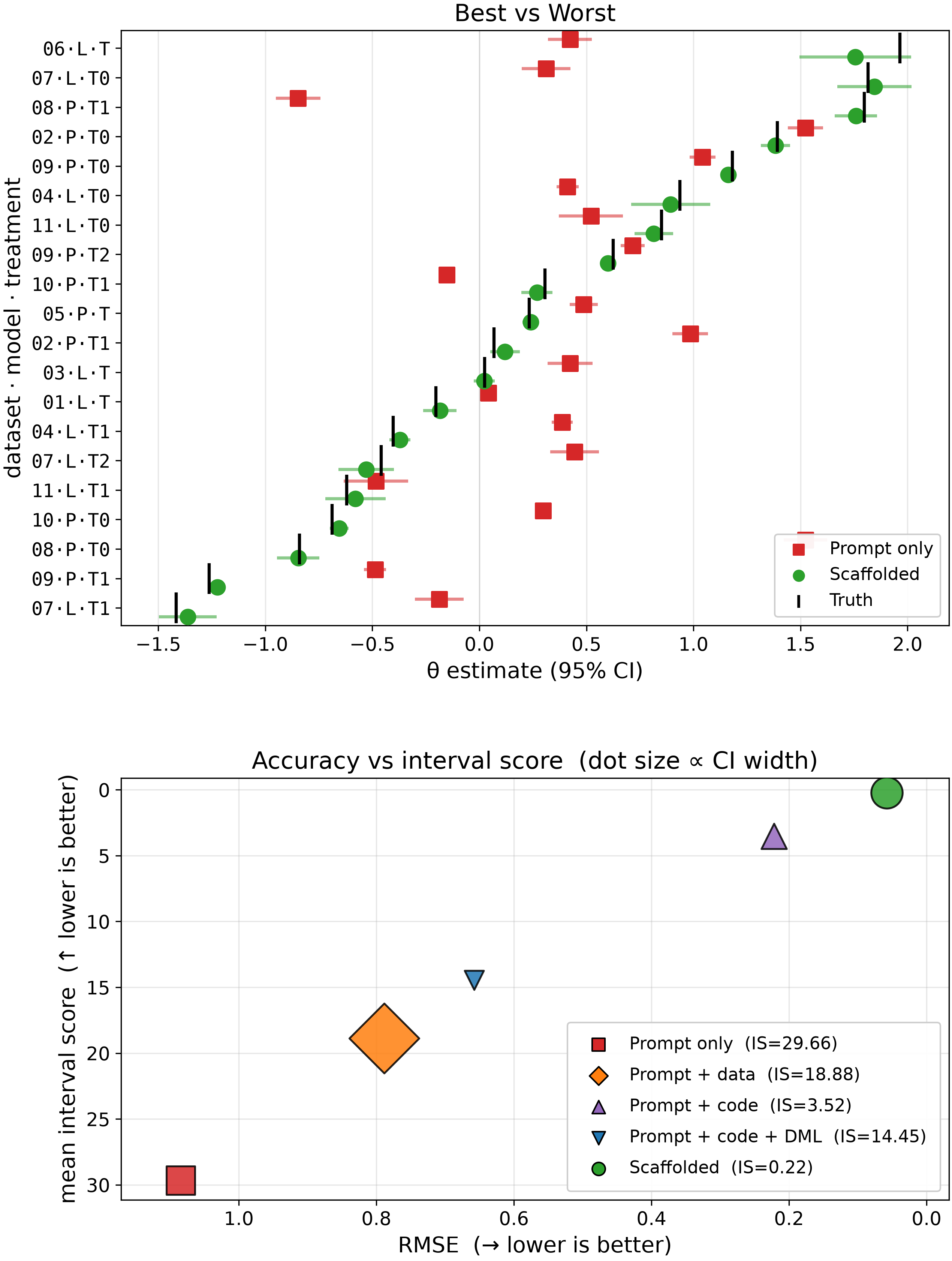}
    \begin{minipage}{.9\columnwidth}
    \caption{PLM $\theta$ estimates (via DoubleML). Top: best vs.\ worst (pure prompting versus \agent~{}'s full scaffolding) against the known coefficients. Bottom: accuracy (RMSE) versus mean interval score across all five layers (dot size proportional to mean CI width).}
    \Description{A stacked figure for the synthetic PLM datasets. The top panel is a forest plot of partially-linear-model coefficient estimates comparing the prompt-only baseline and the scaffolded agent against ground truth, with scaffolded tracking the truth and prompt-only not. The bottom panel is a scatterplot of RMSE versus mean interval score for all five capability-ladder arms, with the scaffolded agent most accurate and best-calibrated.}
    \label{fig:plm}
    \end{minipage}
\end{figure}

Figure~\ref{fig:plm} contrasts pure prompting versus our scaffolding on 20 estimates of $\theta$ from 11 datasets generated from this DGP. As the top panel shows, the model requires scaffolding to produce reliable results; without scaffolding, it produces estimates that are wholly uncorrelated with ground truth. The bottom panel confirms that the scaffolded agent achieves the best accuracy and interval score across the full capability ladder, which again includes tool access, the full dataset, and an explicit instruction to use double machine learning.

\section{Discussion}

Across all three capabilities, we draw four lessons that may transfer to other applied data science agents. First, scaffolding and deterministic tools are necessary for models to succeed on causal inference tasks; without these, even relatively sophisticated models are not yet consistently accurate. Although the underlying models will surely become more capable, equipping them with deterministic tools improves efficiency, inspectability, and reproducibility. Second, when equipped with this scaffolding, agents offer significant value, even to tasks that require substantial human context and judgment. This is because such tasks often involve repetitive and potentially error-prone steps, such as repeating an analysis with different hyperparameters.

Third, human oversight also benefits from an interface. Simple diagnostic tables can flag obvious errors or caveats, and intermediate artifacts, such as plots and notebooks, are critical for inspectability and reproducibility. Fourth, evaluation of causal inference agents needs to combine synthetic benchmarks with process checks. Synthetic datasets are invaluable because they provide ground truth, which is inherently lacking in observational datasets \cite{holland1986statistics}.  Still, these do not replace real-world case studies of the applications on which such agents are ultimately deployed.

\section{Conclusion}
We introduce \agent, a human-in-the-loop agentic workflow for observational causal inference. Built on top of Netflix's pre-LLM OCI infrastructure, our workflow is designed so that agents adhere to comprehensive analysis templates, check design diagnostics, and orchestrate sensitivity analyses. This lessens the human toil of OCI, which can be a highly iterative and exacting process. At the same time, motivated by the complexity and ambiguity of OCI, \agent~enables human oversight by surfacing intermediate artifacts that allow principals to inspect and reproduce each analytic step.  We welcome researchers to use, adapt, and improve \agent, which can be found at \url{https://github.com/Netflix-Skunkworks/oci-agent}.

Using agents for causal inference poses a challenge to practitioners and frontier AI researchers alike: how do we evaluate agents' performance on tasks without ground truth? To meet this challenge, our workflow combines process audits with human inspectability and oversight. Stepping back, we hope this work stimulates more research on agentic evaluation in the absence of ground truth.

\bibliographystyle{ACM-Reference-Format}
\bibliography{verified_references_acmart}


\end{document}